\documentclass[aps,pra,twocolumn,showpacs,twoside]{revtex4}

\usepackage{amsmath,amssymb,latexsym}
\usepackage{mathptmx}

\newcommand{\setN}{\mathbb{N}}

\newcommand{\xor}{\overline{\lor}}
\newcommand{\pitwo}{{}^{\pi}\hspace{-.2em}/\hspace{-.1em}_2}
\newcommand{\NOPA}{\text{NOPA}}
\newcommand{\EPR}{\text{EPR}}
\newcommand{\CHSH}{\text{CHSH}}
\newcommand{\XOR}{\textsc{xor}}
\newcommand{\AND}{\textsc{and}}

\begin{document}

\makeatletter
\pagestyle{myheadings}
\thispagestyle{titlepage}
\def\ps@myheadings{%
    \def\@oddfoot{\hfil 022108-\thepage\hfil}\let\@evenfoot\@oddfoot
    \def\@evenhead{JAN-\AA KE LARSSON\hfil PHYSICAL REVIEW A {\bf 67},
      022108 (2003)}%
    \def\@oddhead{QUBITS FROM NUMBER STATES AND BELL\ldots\hfil
      PHYSICAL REVIEW A {\bf 67}, 022108 (2003)}%
    \let\@mkboth\@gobbletwo \let\sectionmark\@gobble
    \let\subsectionmark\@gobble
    }%
\def\ps@titlepage{%
  \def\@oddfoot{
    {\url{http://link.aps.org/abstract/PRA/v67/e022108}}
       \quad {\bf 67} 022108-\thepage \hfil\copyright2003 The American
      Physical Society}
    \def\@oddhead{\hfil PHYSICAL REVIEW A {\bf 67}, 022108 (2003)\hfil}%
    \let\@mkboth\@gobbletwo
    \let\sectionmark\@gobble
    \let\subsectionmark\@gobble
    }%
\makeatother

\title{Qubits from number states and 
  Bell inequalities for number measurements}


\author{Jan-\AA ke Larsson}\altaffiliation[Present address:
]{Matematiska Institutionen, Link\"opings Universitet, SE-581 83
  Link\"oping, Sweden} \email[E-mail: ]{jalar@mai.liu.se}
\affiliation{MaPhySto, Department of Mathematical Sciences, Aarhus
  University, Ny Munkegade, DK-8000 Aarhus C, Denmark}
\author{(Received 28 August 2002; Published 24 February 2003)}

\begin{abstract}
  Bell inequalities for number measurements are derived via the
  observation that the bits of the number indexing a number state are
  proper qubits.  Violations of these inequalities are obtained from
  the output state of the nondegenerate optical parametric amplifier.
\end{abstract}

\maketitle
\pagestyle{myheadings}

\section{Introduction}

The Bell inequality \cite{Bell64} and its descendants (see e.g.,
Ref.~\cite{CHSH}) are the main tools used when studying the question
whether a bipartite system can be described by a local realist model
or not. In such a model, the properties of the system exist
independently of measurement, and measurements on one subsystem do not
influence the other subsystem. For a proper statement of the exact
properties of a local realist model, see conditions
(\ref{p:realism})--(\ref{p:Result}) below. Under the assumption of
local realism, a bound on the statistics from a bipartite system can
be derived, a Bell inequality, e.g.,~(\ref{eq:Bell}) below. A
bipartite \emph{quantum} system is not described by such a model, and
indeed, the mentioned statistical bound is violated by statistics
obtained from the singlet spin-$\tfrac 12$ quantum state. Thus, that
system \emph{cannot} be described by a local realist model.

The two parts of a bipartite spin-$\tfrac12$ system each have a
quantum description that is (complex) two-dimensional, and this is
appropriate for a violation of the inequalities in
Refs.~\cite{Bell64,CHSH}. There are a number of treatments of
higher-dimensional systems, e.g., Refs.~\cite{KGZMZ,CGLMP}, but here
an attempt will be made to look at an infinite-dimensional system and
its number operator. Some problems that emerge will be discussed and
inequalities suitable for the situation will be derived.  These
inequalities are violated by the state produced in the nondegenerate
optical parametric amplifier (\NOPA).

The interest in these questions stems from the Einstein-Podolsky-Rosen
(\EPR) paradox \cite{EPR}. Perhaps one should comment on the relation
between the continuous-variable, infinite-dimensional quantum system
used in the original \EPR\ paradox, where position ($x$) and momentum
($p$) are used, and the finite-dimensional spin-based approach of
Bohm~\cite{Bohm51}.  Bell \cite{Bell86} has presented a local realist
model for position and momentum measurements on the original \EPR\ 
state, constructed using the Wigner function representation of the
\EPR\ state as a joint probability of the measurement results. The
Wigner function generally has all the properties of a probability
measure except one: it can be negative (a proper probability measure
is always positive). However, in this particular case the Wigner
function is positive, so it can be used as a proper probability
measure. One could be led to think that this implies that the \EPR\ 
state can be described by a local realist model, but this is not the
case: the important thing to note is the statement \emph{``for
  position and momentum measurements.''}  That is, nothing is said
about \emph{other} measurements; the quantum state contains more than
just information about position and/or momentum. In fact, if one
instead uses measurements of parity, one can interpret the Wigner
function as a correlation function for these parity measurements, and
then regularized \EPR\ states \emph{are} nonlocal \cite{BanaWodk}.

Furthermore, in Ref.~\cite{CPHZ}, pseudo-spin operators based on
parity are introduced, and from them a violation of the
Clauser-Horne-Shimony-Holt (\CHSH) inequality \cite{CHSH} is derived
using the $|\NOPA\rangle$ state. The steps will be briefly reiterated
here as they will be important in what follows. The pseudo-spin
operators are
  \begin{equation}
    \label{eq:s1}
    \begin{split}
      s_z&=\sum_{n=0}^{\infty}(-1)^n|n\rangle\langle n|,\\
      s_+&=(s_-)^{\dag}=\sum_{n=0}^{\infty}|2n\rangle\langle2n+1|,\\
      s_x&=s_++s_-,\quad s_y=-i(s_+-s_-),
    \end{split}
  \end{equation}
  and these satisfy the usual commutation relations
\begin{equation}
  \label{eq:1}
  [s_z,s_{\pm}]=\pm2s_{\pm},\quad [s_+,s_-]=s_z.
\end{equation}
Please observe the change of sign convention in Eq.~(\ref{eq:s1}) from
that of Ref.~\cite{CPHZ}. Now, the pseudo-spin operator
$\mathbf{\hat{s}}=(s_x,s_y,s_z)$ in a sense corresponds to the spin
operator $\mathbf{\hat{\pmb{\sigma}}}$ for a normal spin-$\tfrac12$
system. In the general case, a measurement of pseudospin can be made
along a certain ``direction'' $\mathbf{a}$ by using the operator
$\mathbf{a}\cdot\mathbf{\hat{s}}$, but for our purposes, planar
variation is sufficient, so let us use the notation
\begin{equation}
  s_{\theta}=\cos(\theta)s_z+\sin(\theta)s_x.
\end{equation}
The two parts of our bipartite system can be subjected to individual
measurements of the above type, and we will use the shorthand notation
$s_{\alpha}s'_{\beta}$ to denote $s_{\alpha}\otimes s_{\beta}$ below.

In our bipartite system we will use the $|\NOPA\rangle$ state as our
entangled state
\begin{equation}
  |\NOPA\rangle
  =\frac{1}{\cosh r}
  \sum_{n=0}^{\infty}\tanh^n r|n\rangle\otimes|n\rangle.
\end{equation}
The parameter $r$ is usually referred to as the \emph{squeezing
  parameter}, and is a measure of the amount of squeezing in the
system. This state is sometimes referred to as a \emph{regularized}
\EPR\ state, and at infinite squeezing ($r\rightarrow\infty$), the state
approaches the idealized state used in the original EPR paper
\cite{EPR}. A simple calculation yields \cite{CPHZ}
\begin{equation}
\begin{split}
  \langle\NOPA|s_zs'_z|\NOPA\rangle
  &=1,\\
  \langle\NOPA|s_xs'_x|\NOPA\rangle
  &=\tanh 2r=:K.
\end{split}\label{eq:K}
\end{equation}
$K$ is a (strictly increasing) function of $r$, $K=0$ when $r=0$ and
$K\rightarrow 1$ when $r\rightarrow\infty$; consequently $K$ can
equally well be used as a measure of the amount of squeezing in this
state. Finally, we have
\begin{equation}
  \langle\NOPA|s_{\alpha}s'_{\beta}|\NOPA\rangle
  =\cos{\alpha}\cos{\beta}+K\sin{\alpha}\sin{\beta}.
\end{equation}

The results of the individual measurements $s_{\alpha}\otimes I$ and
$I \otimes s_{\beta}$ will be denoted by $S_{\alpha}$ and
$S'_{\beta}$. That is, these are the classical $\pm1$ values
registered from measurement, e.g., written down on a piece of paper or
similar.  The question is now if these results can be described under
the assumption of local realism:
\begin{enumerate}
\renewcommand{\theenumi}{\roman{enumi}}
\renewcommand{\labelenumi}{(\theenumi)}
\item \label{p:realism} \emph{Realism,} There is a classical
  probabilistic model where the results depend on a ``hidden
  variable'' $\lambda$, i.e.,
  \begin{equation}
    \begin{split}
      S_{\alpha}&=S_{\alpha}(\lambda),\\
      S'_{\beta}&=S'_{\beta}(\lambda).
    \end{split} \label{eq:2}
  \end{equation}
\item \label{p:locality} \emph{Locality,} The model is local, such
  that measurement settings at one subsystem does not affect the other
  subsystem, i.e.,
  \begin{equation}
    \begin{split}
      S_{\alpha}(\lambda)&\text{ is independent of }\beta,\\
      S'_{\beta}(\lambda)&\text{ is independent of }\alpha.
    \end{split}
  \end{equation}
\item \label{p:Result} \emph{Result restriction.} The measurement
  results are restricted in size,
  \begin{equation}
    |S_{\alpha}(\lambda)|\le 1,\quad |S'_{\beta}(\lambda)|\le 1.
  \end{equation}
\end{enumerate}
When this is the case, we have the \CHSH\ inequality \cite{CHSH}
\begin{equation}
  \big|E(S_{\alpha}S'_{\gamma})
  +E(S_{\alpha}S'_{\delta})\big|
  +\big|E(S_{\beta}S'_{\gamma})
  -E(S_{\beta}S'_{\delta})\big|
  \leq2.\label{eq:Bell}
\end{equation}
However, using the $|\NOPA\rangle$ state (in shorthand notation), we have
\begin{equation}
  \begin{split}
    &\big|\langle s_{0}s'_{\gamma}\rangle
    +\langle s_{0}s'_{-\gamma}\rangle\big|
    +\big|\langle s_{\pitwo}s'_{\gamma}\rangle
    -\langle s_{\pitwo}s'_{-\gamma}\rangle\big| \\ 
    &\qquad\qquad\qquad\qquad\qquad\quad
    =2(\left|\cos\gamma\right|+K\left|\sin\gamma\right|).
  \end{split}\label{eq:qm_ss}
\end{equation}
With e.g., $\gamma=\arctan K$, the maximum of the right-hand side is
obtained at $2\sqrt{1+K^2}>2$.  The conclusion is that
(\ref{p:realism}) or (\ref{p:locality}) (or both) must fail, since
(\ref{p:Result}) always holds in this setting.

In the ideal case, there is a violation at any nonzero squeezing, but
in a noisy setting the violation will be lowered by the noise, so that
a certain lowest squeezing will be required. Note that when $K=1$, the
violation will be as large as that generated by the singlet state in
the original setting. This corresponds to an infinite squeezing
parameter $r$, i.e., the original \EPR\ state \cite{EPR}, but
unfortunately infinite squeezing cannot be achieved in practice. Note
that the angle $\gamma$, at which there is maximum violation, depends
on the squeezing.

The detector-efficiency problem \cite{Jalar98a} is less of an issue
here than in the usual Bell inequality because e.g., lost photons in
an optical implementation will not imply that experimental runs are
dropped from the statistics. Instead such losses will change the
measured parity, introducing noise in the statistics and leading to a
situation similar to that of the ion-trap experiment of Rowe
\emph{et~al.}\ \cite{Rowe01}, where also ``dark'' events lead to
increased noise in the experimental data. Thus, we do not need to use
auxiliary assumptions such as the no-enhancement assumption
\cite{ClauHorn}. One of the motivations of Refs.~\cite{BanaWodk,CPHZ}
and the present paper is to derive a violation of local realism from
continuous-variable systems using only the assumptions (i)--(iii)
above. A different philosophy is used in some previous proposals for
continuous-variable Bell inequalities, where additional assumptions
are necessary and in some cases built into the formalism (see e.g.,
Refs.~\cite{GPY,TWC91,Santos92,MR93,YS97,RMP00,HR02}). It should be
noted that usage of these assumptions usually lead to simpler
experimental implementations than is expected from the present
treatment.

\section{An infinite commuting hierarchy}

To extend this, let us use a different assignment of parity than the
usual one. As an example, group the number states two-by-two, assign
``even'' parity to the number states in the first group and ``odd''
parity to the next group, and so on. Above, we had
\begin{equation}
  \label{eq:sz}
    s_z=
    +|0\rangle\langle 0|-|1\rangle\langle 1|
    +|2\rangle\langle 2|-|3\rangle\langle 3|
    +\cdots,
\end{equation}
but now, a similar expression would be
\begin{equation}
  \label{eq:sz2}
  \begin{split}
    s_{z,2}=&+\Big(|0\rangle\langle 0|+|1\rangle\langle 1|\Big)
    -\Big(|2\rangle\langle 2|+|3\rangle\langle 3|\Big)\\
    &+\Big(|4\rangle\langle 4|+|5\rangle\langle 5|\Big)
    -\Big(|6\rangle\langle 6|+|7\rangle\langle 7|\Big)
    +\cdots
  \end{split}
\end{equation}
Adjusting the expressions for $s_{\pm}$ in eq.~(\ref{eq:s1}) for this
case, we obtain
\begin{align}
  \label{eq:s2}
  s_{z,2}&=\sum_{n=0}^{\infty}(-1)^n
  \Big(|2n\rangle\langle 2n|+|2n+1\rangle\langle 2n+1|\Big)\notag\\
  s_{+,2}&=(s_{-,2})^{\dag}=\sum_{n=0}^{\infty}
  \Big[|4n\rangle\langle4n+2|+|4n+1\rangle\langle4n+3|\Big],\notag\\
  s_{x,2}&=s_{+,2}+s_{-,2},\quad s_{y,2}=-i(s_{+,2}-s_{-,2}).
\end{align}
Also in this case, it is easy to check that there is a correspondence
to a spin-$\tfrac12$ system.

Extending this to arbitrary $d$-by-$d$ grouping is equally simple,
\begin{align}
  \label{eq:sd}
  s_{z,d}&=\sum_{n=0}^{\infty}(-1)^n
  \sum_{k=0}^{d-1} |dn+k\rangle\langle dn+k|,\notag\\
  s_{+,d}&=(s_{-,d})^{\dag}
  =\sum_{n=0}^{\infty}
  \sum_{k=0}^{d-1} |2dn+k\rangle\langle 2dn+k+d|,\notag\\
  s_{x,d}&=s_{+,d}+s_{-,d},\quad s_{y,d}=-i(s_{+,d}-s_{-,d}).
\end{align}
For example,
\begin{equation}
  \label{eq:sz3}
  \begin{split}
    s_{z,3}=&+\Big(|0\rangle\langle 0|+|1\rangle\langle 1|
    +|2\rangle\langle 2|\Big)\\
    &-\Big(|3\rangle\langle 3|+|4\rangle\langle 4|
    +|5\rangle\langle 5|\Big)\\
    &+\Big(|6\rangle\langle 6|+|7\rangle\langle 7|
    +|8\rangle\langle 8|\Big) +\cdots,
  \end{split}
\end{equation}
while the $d=3$ spin step would be
\begin{equation}
  \begin{split}
    s_{+,3}&=|0\rangle\langle 3|+|1\rangle\langle 4|
    +|2\rangle\langle 5|\\
    &\quad+|6\rangle\langle 9|+|7\rangle\langle 10|
    +|8\rangle\langle 11|+\ldots
  \end{split}
\end{equation}
Another simple calculation yields
\begin{equation}
  \begin{split}
    \langle\NOPA|s_{z,d}s'_{z,d}|\NOPA\rangle&=1,\\
    \langle\NOPA|s_{x,d}s'_{x,d}|\NOPA\rangle&
    =\frac{2\tanh^d r}{1+\tanh^{2d} r}=:K_d.
    \label{eq:K_d}
  \end{split}
\end{equation}
Note that if $0<r<\infty$, $K_d$ decreases when $d$ increases. In a
similar fashion as before,
\begin{equation}
   \langle\NOPA|s_{\alpha,d}s'_{\beta,d}|\NOPA\rangle
  =\cos{\alpha}\cos{\beta}+K_d\sin{\alpha}\sin{\beta}.
  \label{eq:correlation}
\end{equation}
With the same notational conventions as above, we have under local
realism ((\ref{p:realism})--(\ref{p:Result})), that
\begin{equation}
  \begin{split}
    &\big|E(S_{\alpha,d}S'_{\gamma,d})
    +E(S_{\alpha,d}S'_{\delta,d})\big|\\
    &\qquad+\big|E(S_{\beta,d}S'_{\gamma,d})
    -E(S_{\beta,d}S'_{\delta,d})\big|
    \leq2,
  \end{split}
  \label{eq:Bell_d}
\end{equation}
and similarly to Eq.~(\ref{eq:qm_ss}),
\begin{equation}
  \begin{split}
    &\big|\langle s_{0,d}s'_{\gamma,d}\rangle
    +\langle s_{0,d}s'_{-\gamma,d}\rangle\big|
    +\big|\langle s_{\pitwo,d}s'_{\gamma,d}\rangle
    -\langle s_{\pitwo,d}s'_{-\gamma,d}\rangle\big| \\ 
    &\qquad\quad
    =2(\left|\cos\gamma\right|+K_d\left|\sin\gamma\right|).
  \end{split}\label{eq:qm_ssd}
\end{equation}
Again for nonzero squeezing a violation of Ineq.~(\ref{eq:Bell_d}) is
obtained at any positive $d$, but it decreases as $d$ grows. Here,
maximum-violation $\gamma$ depends both on the squeezing and on $d$.
 
It is not generally true that the pseudospin operators commute for
different $d$. For example, the operators for $d=2$ and $d=3$ do not
commute; while it is true that $[s_{z,2},s_{z,3}]=0$, we have
$[s_{x,2},s_{x,3}]\neq0$. However, an important case when they
\emph{do} commute is when the $d$'s are related by multiplication by
an even number. This is easily verified by inspection in the relation
between $d=1$ as in Eq.~(\ref{eq:s1}) and $d'=2$ as in
Eq.~(\ref{eq:s2}), and can also be extended to any situation where
$d'=2^k d$; the spin operators $s_{\pm,d}$ then only exchange number
states entirely within the groups defined by $s_{z,d'}$ and
$s_{\pm,d'}$, and $s_{z,d}$ performs the same sign change within each
such group. We obtain an infinite commuting hierarchy of spin systems,
\begin{equation}
  \{s_{z,2^k},s_{-,2^k}=(s_{+,2^k})^{\dag}\}_{k\in\setN}.
\end{equation}
Using the bipartite $|\NOPA\rangle$ state, there is a simultaneous,
separate violation of a Bell inequality within each spin system in
this hierarchy.

\section{New number operators?}

Interestingly, the $s_{z,2^k}$ operator corresponds to measurement of
the \emph{bits in the binary representation of $n$}, mapping a
bit-value 0 into the parity value $+1$ and a bit-value 1 into parity
$-1$. This is readily seen in Eqs.~(\ref{eq:sz}) and (\ref{eq:sz2})
and can be shown generally using Eq.~(\ref{eq:sd}). The correspondence
is
\begin{equation}
  s_{z,2^k}=1-2b_k,\label{eq:b-s}
\end{equation}
wherein $b_k$ is the $k$th bit of $n$. Obtaining number from parity is
also possible,
\begin{equation}
  n=\sum_{k=0}^{\infty}2^kb_k=\sum_{k=0}^{\infty}2^k\frac{1-s_{z,2^k}}{2}.
\end{equation}
Here, it is \emph{very important} that the different $s_{z,2^k}$
commute. The above construction is, more or less, the observation that
the bits of the number representation are qubits in the usual sense of
the word, and we can see above that in the $|\NOPA\rangle$ state,
there is indeed entanglement of the qubits in the number
representation.

The bit correspondence of the $s_{z,2^k}$ operator to the number
operator $n$ (henceforth referred to as $n_z$) hints at similar
constructions of number operators $n_x$ and $n_y$ corresponding to
$s_{x,2^k}$ and $s_{y,2^k}$, for example,
\begin{equation}
  n_x=\sum_{k=0}^{\infty}2^k\frac{1-s_{x,2^k}}{2}.
\end{equation}

Unfortunately, this construction is problematic, for example, the
eigenstates of $n_x$ and $n_y$ will not be normal states; let us
determine $|0_x\rangle$. All the $n_x$ bits are zero which corresponds
to the eigenvalue $+1$ of all $s_{x,2^k}$. From Eq.~(\ref{eq:s1}) an
eigenstate $|\psi\rangle$ of $s_{x,1}$ with the eigenvalue $+1$ will
have the property
\begin{equation}
  \langle0_z|\psi\rangle = \langle1_z|\psi\rangle,     
\end{equation}
and similarly, from Eq.~(\ref{eq:s2}) an eigenstate $|\varphi\rangle$
of $s_{x,2}$ with the eigenvalue $+1$ will follow
\begin{equation}
  \begin{split}
    \langle0_z|\varphi\rangle &= \langle2_z|\varphi\rangle,\\
    \langle1_z|\varphi\rangle &= \langle3_z|\varphi\rangle.
  \end{split}
\end{equation}
Continuing (infinitely), a simultaneous eigenstate of $s_{x,2^k}$ with
the eigenvalue $+1$ for all $k$'s will have all coefficients equal in
the $n_z$ basis, i.e., of the form
\begin{equation}
  C\sum_{n=0}^{\infty}|n_z\rangle.
\end{equation}
This is not a normal state; it has infinite energy. Nevertheless, it
has a bit value of 0 at all positions when measuring $n_x$, so
choosing $|0_x\rangle$ so that $\langle0_z|0_x\rangle=1$, we have (with
a certain abuse of notation \footnote{In a situation like this, one
  should really deal with states in the form of functionals on
  operators and weak-* convergence, but perhaps this would lead us
  astray in our search for a Bell inequality.})
\begin{equation}
  |0_x\rangle=\sum_{n=0}^{\infty}|n_z\rangle.
\end{equation}
 
More generally, eigenstates to the eigenvalues $\pm1$ of the spin
operator $s_{x,d}$ (see Eq.~(\ref{eq:sd})) have the properties
\begin{equation}
  \label{eq:eigstate}
  \begin{split}
    \langle 0_z|\varphi\rangle
    &= 
    \pm\langle d_z|\varphi\rangle,\\
    \langle 1_z|\varphi\rangle
    &= 
    \pm\langle(d+1)_z|\varphi\rangle,\\
    &\;\;\vdots\\
    \langle(d-1)_z|\varphi\rangle
    &= \pm\langle(2d-1)_z|\varphi\rangle,
  \end{split}
\end{equation}
where the eigenvalue in question decides what sign the
right-hand sides have. The vector $|m_x\rangle$ is a simultaneous
eigenvector to all $s_{x,2^k}$, and by choosing it so that
\begin{equation}
  \label{eq:5}
  \langle0_z|m_x\rangle=1, 
\end{equation}
we obtain [put $d=2^k$ in Eq.~(\ref{eq:eigstate})]
\begin{equation}
  \label{eq:m_x}
  \begin{split}
    \langle (2^k)_z|m_x\rangle &= \pm1,\\
    \langle (2^k+1)_z|m_x\rangle
    &= \langle (2^k)_z|m_x\rangle\langle 1_z|m_x\rangle,\\
    &\;\;\vdots\\
    \langle(2^{k+1}-1)_z|m_x\rangle &= \langle
    (2^k)_z|m_x\rangle\langle(2^k-1)_z|m_x\rangle.
  \end{split}
\end{equation}
To determine the sign above, first note that
\begin{equation}
    \langle1_z|m_x\rangle=
    \begin{cases}
      +1, & m\text{ even}\\
      -1, & m\text{ odd}.
    \end{cases}
\end{equation}
This is because $|m_x\rangle$ is an eigenvector to $s_{x,1}$ with
eigenvalue $+1$ if the lowest bit of $m$ is zero ($m$ is even), and
with eigenvalue $-1$ if the lowest bit of $m$ is one ($m$ is odd). In
other words, $\langle1_z|m_x\rangle =(-1)^{m\land 1}$, where $\land$
denotes the bitwise \AND operation on the two numbers. Continuing, we
have
\begin{equation}
    \langle(2^k)_z|m_x\rangle=
    \begin{cases}
      +1, & m \land 2^k=0\\
      -1, & m \land 2^k=2^k.
    \end{cases}
\end{equation}
And finally, expanding $n$ binary,
\begin{equation}
  \begin{split}
    &\langle n_z|m_x\rangle=
    \langle\Big(\sum_{k=1}^K b_k2^k\Big)_z|m_x\rangle\\
    &\quad=\langle (b_K2^K)_z|m_x\rangle
    \langle\Big(\sum_{k=1}^{K-1} b_k2^k\Big)_z|m_x\rangle\\
    &\quad =\ldots =\prod_{k=1}^K \langle(b_k2^k)_z|m_x\rangle.
  \end{split}
\end{equation}
In the right-hand side above, there will be a $-1$ factor each time
$m\land b_k2^k$ is nonzero. By simply counting the number of bits that
are set in $m\land n$ [and denoting the result ${\cal N}(m\land n)$], we
arrive at
\begin{equation}
  |m_x\rangle=\sum_{n=0}^{\infty}
  (-1)^{{\cal N}(m\land n)}|n_z\rangle.
\end{equation}
In the same manner, one can deduce
\begin{equation}
  |m_y\rangle=\sum_{n=0}^{\infty}
  (-1)^{{\cal N}(m\land n)}i^{{\cal N}(n)}|n_z\rangle.
\end{equation}
The connection between the eigenstates of $n_x$ and $n_y$ is more
problematic since all of these are non-normal, and this relation will
not be reproduced here \cite{endnote21}.

Unfortunately, the non-normal eigenstates of $n_x$ implies that when
measuring $n_x$ the state would be projected onto a non-normal state
(for a finite measurement result, using von Neumann measurement
theory). Such a measurement would generate an infinite-energy output
quantum state, which seems to be a serious deficit in this approach.

Conversely, consider a measurement of $n_x$ on the vacuum. The state
$|0_z\rangle$ (the vacuum) is an eigenstate to the original number
operator $n_z$ that has zeros at all bit values, corresponding to the
eigenvalue $+1$ of $s_{z,2^k}$ for all $k$'s. The properties of a spin
system tell us that it is to be expected that a measurement of any
$s_{x,2^k}$ will yield equally probable results $\pm1$. Thus, the bits
of $n_x$ will be evenly distributed, and the probability of getting a
finite result from a measurement of $n_x$ on the vacuum is zero. 

This is actually true for any finite-energy state $|\psi\rangle$: as
$k$ grows in $s_{z,2^k}$, we are looking at higher and higher bits of
$n_z$. Since the state has finite energy, the probability of a bit
being set will tend to zero as $k$ grows, and equivalently the
probability of the result $+1$ when measuring $s_{z,2^k}$ will tend to
zero as $k$ grows. This in turn means that the probability of the
result $+1$ when measuring $s_{x,2^k}$ will tend to $\tfrac 12$, and
equivalently that the probability of bit $k$ being unset in a
measurement of $n_x$ will tend to $\tfrac 12$. The probability may
never reach $\tfrac12$, but we know that whenever $k$ exceeds some
(large) $K$, this probability will be less than $\tfrac23$. Now, the
probability of getting a result bounded by $2^k$ is
\begin{equation}
  \label{eq:7}
  \begin{split}
    \langle\psi|\chi_{n_x\le2^k}|\psi\rangle
    <\prod_{l=k+1}^{k+L}\langle\psi|\chi_{n_x\land 2^l=0}|\psi\rangle
    <\big(\tfrac23\big)^L \rightarrow0,&\\ L\rightarrow \infty,&
  \end{split}
\end{equation}
Here, $\chi_{A}$ denotes the projector onto the subspace where the
property $A$ holds. Ineq.~(\ref{eq:7}) is valid for any $k$ (larger
than $K$) and thus, for any finite-energy state, measurement of $n_x$
(almost) never yields a finite result.

The above-mentioned problems are two sides of the same coin: if a
measurement of $n_x$ yields a finite value, the output state from the
measurement process has infinite energy; conversely, if a
finite-energy state is input into a measurement of $n_x$, the result
is always infinite. It does seem very problematic to construct a Bell
inequality for this type of number measurements.  In fact, the
original number operator itself is an unbounded operator and we want
to derive a statistical bound for it (our desired Bell inequality).
One way around these problems is to truncate the operators at the
$d$th bit, and we will use, e.g.,
\begin{equation}
  n_{z,d}=\sum_{k=0}^{d-1}2^k\frac{1-s_{z,2^k}}{2}.
\end{equation}
These operators all yield finite measurement results and finite-energy
output states (with a finite-energy input state), and enables the
construction of a Bell inequality.

\section{A Bell inequality for number measurements}

We have now constructed an infinite commuting hierarchy of pseudospin
systems, each violating a Bell inequality. We have also established
that the $s_{z,2^k}$ operator of each pseudospin system corresponds
to one of the bits of the number operator $n$($=n_z$), and more
interestingly, each such bit is a qubit in the standard sense of the
word. Measurement results can now be represented equally well in the
language of spin ($\pm1$) as in the language of bits
[$\frac12-(\pm\frac12)$], and we should be able to rewrite our Bell
inequalities~(\ref{eq:Bell}) and~(\ref{eq:Bell_d}) in the language of
bits rather than spin.

To complete this rewrite, one question remains: we multiply
spin-measurement results ($SS'$) in our Bell inequalities but what
function $f(B,B')$ does this correspond to in the bit language?
Listing classical bit values, the corresponding pseudospin values,
their product, and the final corresponding bit value in a table, we
get
\begin{center}
  \begin{tabular}{c c c c|c c}
    $B$&$B'$&$S$&$S'$&$SS'$&$f(B,B')$\\
    \hline
    0&0&$+1$&$+1$&$+1$&0\phantom{.}\\
    0&1&$+1$&$-1$&$-1$&1\phantom{.}\\
    1&0&$-1$&$+1$&$-1$&1\phantom{.}\\
    1&1&$-1$&$-1$&$+1$&0.
  \end{tabular}
\end{center}
Interestingly, $f$ proves to be exclusive-or (\XOR, $\xor$; our sign
convention was useful here). This is natural, since the multiplication
of our spin values is conveniently interpreted as a test whether the
spin values are equal or not; in bit language, such a test is provided
by \XOR.  Moreover,
\begin{equation}
  SS'=(1-2B)(1-2B')=1-2B\xor B'.
\end{equation}
Under conditions (\ref{p:realism})--(\ref{p:Result}) of local realism,
and using
\begin{equation}
  \label{eq:6}
  B_{\theta,k}={\tfrac 12}(1-S_{\theta,2^k})
\end{equation}
we obtain the following directly from Eq.~(\ref{eq:Bell_d}):
\begin{equation}
  \begin{split}
    &\big|E(B_{\alpha,k}\xor B'_{\gamma,k})
    +E(B_{\alpha,k}\xor B'_{\delta,k})-1\big|
    \\&\quad
    +\big|E(B_{\beta,k}\xor B'_{\gamma,k})
    -E(B_{\beta,k}\xor B'_{\delta,k})\big|
    \\&
    =\tfrac{1}{2}\big|E(S_{\alpha,2^k}S'_{\gamma,2^k})
    +E(S_{\alpha,2^k}S'_{\delta,2^k})\big|
    \\&\phantom{=}\quad
    +\tfrac{1}{2}\big|E(S_{\beta,2^k}S'_{\gamma,2^k})
    -E(S_{\beta,2^k}S'_{\delta,2^k})
    \big|\label{eq:Bell_B}
    \leq1.
  \end{split}
\end{equation}
A corollary is the more familiar-looking inequality,
\begin{equation}
  \begin{split}
    &\big|E(B_{\alpha,k}\xor B'_{\gamma,k})
    +E(B_{\alpha,k}\xor B'_{\delta,k})\big|
    \\&\quad+\big|E(B_{\beta,k}\xor B'_{\gamma,k}) -E(B_{\beta,k}\xor
    B'_{\delta,k})\big| \leq2.
  \end{split}
\end{equation}
but Eq.~(\ref{eq:Bell_B}) is tighter and will be used below.
  
We would now like to check for a violation of the above inequality
from quantum mechanics, preferably from the previously mentioned
$|\NOPA\rangle$ state. The multiplication of pseudospin
\emph{operators} (quantum-mechanical) translates into a ``quantum
\XOR''
\begin{equation}
  ss'=(1-2b)(1-2b')=1-2b\xor b',
\end{equation}
so that 
\begin{equation}
  b\xor b'=b+b'-2bb'.
\end{equation}
This has all the properties expected from an \XOR operation, and is a
noncommutative operation, e.g.,
\begin{equation}
  \begin{split}
    b_x\xor b_z-b_z\xor b_x
    &=\frac{1-s_xs_z}2- \frac{1-s_zs_x}2
    =\frac{[s_z,s_x]}2\\
    &=is_y=i(1-2b_y).
  \end{split}
\end{equation}
Using the above, it is easy to check that 
\begin{equation}
  \begin{split}
    &\big|\langle b_{0,d}\xor b'_{\gamma,d}\rangle
    +\langle b_{0,d}\xor b'_{-\gamma,d}\rangle-1\big|\\
    &\qquad
    +\big|\langle b_{\pitwo,d}\xor b'_{\gamma,d}\rangle
    -\langle b_{\pitwo,d}\xor b'_{-\gamma,d}\rangle\big| \\ 
    &\qquad\qquad\qquad\qquad
    =\left|\cos\gamma\right|+K_d\left|\sin\gamma\right|.
  \end{split}\label{eq:qm_bbd}
\end{equation}
We again have a violation, although both the bound and the violation
are a factor of $\tfrac12$ less than in Ineqs.~(\ref{eq:Bell_d}) and
(\ref{eq:qm_ssd}). This makes the absolute quantum-mechanical
violation of Ineq.~(\ref{eq:Bell_B}) less than that of
Ineq.~(\ref{eq:Bell_d}), but the sensitivity to experimental problems
is the same; we have simply represented the data differently.

Until now we have treated the pseudospin systems as individual
systems, measuring and comparing them one-by-one.  Let us now do a
joint treatment of the first $d$ bits by using the measurement results
$N_{z,d}$ and so on, written in binary form, for example,
\begin{equation}
  \label{eq:3}
  N_{z,d}=\sum_{k=0}^{d-1}2^kB_{z,k}.
\end{equation}
The truncation to $d$ bits is useful to avoid the aforementioned
problems of the number operators, but will also be necessary to give a
useful bound below. We now perform a comparison of the individual bits
in each (truncated) number by using bitwise \XOR,
\begin{equation}
  \label{eq:8}
  \begin{split}
    N_{\alpha,d}\xor N'_{\gamma,d}
    &=(\sum_{k=0}^{d-1}2^kB_{\alpha,k})\xor
    (\sum_{l=0}^{d-1}2^lB'_{\gamma,l})\\
    &=\sum_{k=0}^{d-1}2^k(B_{\alpha,k}\xor B'_{\gamma,k}).
  \end{split}
\end{equation}
For each individual bit, we have the Bell inequality~(\ref{eq:Bell_B})
violated by the quantum-mechanical expression~(\ref{eq:qm_bbd}), and
joining these together, we have
\begin{widetext}
\begin{small}
  \begin{align}
    &\big|E(N_{\alpha,d}\xor N'_{\gamma,d})
      +E(N_{\alpha,d}\xor N'_{\delta,d})-(2^d-1)\big|
      +\big|E(N_{\beta,d}\xor N'_{\gamma,d})
      -E(N_{\beta,d}\xor N'_{\delta,d})\big|
      \notag\\&
      =\Big|
      E\Bigg(\sum_{k=0}^{d-1}2^k(B_{\alpha,k}\xor B'_{\gamma,k})\Bigg)
      +E\Bigg(\sum_{k=0}^{d-1}2^k(B_{\alpha,k}\xor B'_{\delta,k})\Bigg)
      -\sum_{k=0}^{d-1}2^k\Big|
      +\Big|
      E\Bigg(\sum_{k=0}^{d-1}2^k(B_{\beta,k}\xor B'_{\gamma,k})\Bigg)
      -E\Bigg(\sum_{k=0}^{d-1}2^k(B_{\beta,k}\xor B'_{\delta,k})\Bigg)
      \Big|\notag\\&
      \leq\sum_{k=0}^{d-1}2^k\Big[\big|
      E(B_{\alpha,k}\xor B'_{\gamma,k})
      +E(B_{\alpha,k}\xor B'_{\delta,k})-1\big|
      +\big|E(B_{\beta,k}\xor B'_{\gamma,k})
      -E(B_{\beta,k}\xor B'_{\delta,k})\big|\Big]
      \leq\sum_{k=0}^{d-1}2^k=2^d-1.
      \label{eq:Bell_N}
  \end{align}
\end{small}
\end{widetext}
Again we have a more familiar-looking corollary
\begin{equation}
  \begin{split}
    &\big|E(N_{\alpha,d}\xor N'_{\gamma,d})
    +E(N_{\alpha,d}\xor N'_{\delta,d})\big|
    \\&\, +\big|E(N_{\beta,d}\xor N'_{\gamma,d}) -E(N_{\beta,d}\xor
    N'_{\delta,d})\big| \leq2(2^d-1),
  \end{split}
\end{equation}
and again the tighter inequality (\ref{eq:Bell_N}) will be retained;
it is our desired Bell inequality for number measurements.

A violation of Ineq.~(\ref{eq:Bell_N}) is obtained using the
$|\NOPA\rangle$ state, for which
\begin{equation}
  \begin{split}
    &\big|\langle n_{0,d}\xor n'_{\gamma,d}\rangle +\langle n_{0,d}\xor
    n'_{-\gamma,d}\rangle-(2^d-1)\big|\\
    &\quad\quad\quad
    +\big|\langle n_{\pitwo,d}\xor n'_{\gamma,d}\rangle
    -\langle n_{\pitwo,d}\xor n'_{-\gamma,d}\rangle\big|\\
    &\qquad\qquad\quad
    =(2^d-1)\left|\cos\gamma\right|+\sum_{k=0}^{d-1}2^k
    K_{2^k}\left|\sin\gamma\right|,\label{eq:qm_nn}
  \end{split}
\end{equation}
and with $\gamma$ chosen properly, the maximum violation is obtained
as
\begin{equation}
  \sqrt{(2^d-1)^2+\Big(
    \sum_{k=0}^{d-1}2^kK_{2^k}\Big)^2}
  >2^d-1.
\end{equation}
At infinite squeezing ($K_{2^k}=1$ for all $k$) we have
$\sqrt2(2^d-1)$, which gives the largest possible violation of
Ineq.~(\ref{eq:Bell_N}) \footnote{A technical note is that the first
  inequality of Eq.~(\ref{eq:Bell_N}) cannot be violated by quantum
  mechanics; while the triangle inequality is used there too (it is
  also the basis of the Bell inequality), it is used to split
  \emph{commuting} measurements rather than noncommuting ones.}.

One reason for using a bitwise operation here instead of the usual
multiplication is the above-mentioned correspondence to multiplication
of pseudospins, but there is another, perhaps less evident reason. It
is clear that for the $|\NOPA\rangle$ state, the result of a
measurement of $n_z$ is the same as that of $n_z'$ in the ideal case.
This implies that the measurement results of the bits $b_k$ and $b_k'$
are also the same. In a noisy environment, the correlation would be
high. However, different bits are not correlated in the same manner,
e.g., measurement results from $b_k$ and $b'_{k+1}$ do not enjoy the
same degree of correlation. In fact, even measuring on one part of the
bipartite $|\NOPA\rangle$ state, knowledge of the value of one of the
bits (of $n_z$, say) gives no statistical information on the value of
another. When calculating a product, e.g.,
\begin{equation}
    N_{z,2}N'_{z,2}=B_0B'_0+4B_1B'_1+2(B_0B'_1+B_1B'_0),\label{eq:nn}
\end{equation}
it is easy to see that there are terms that mix \emph{different} bits
of the two numbers. It is of course possible to derive a Bell
inequality for this expression as well, and the first two terms
correspond to a weighted bitwise \AND, similar to the bitwise \XOR
used in this paper. Unfortunately, the parenthesis at the end performs
a bitwise \AND on different bits of the two numbers. Thus, the last
parenthesis yields no usable information, but only additional noise.
It would be possible to get around the noise addition by using a state
in which knowledge of one bit implies knowledge of another, but this
will in effect reduce the available dimensionality of the system; we
would, in bit notation, want to use a state close to
$|00\rangle\otimes|00\rangle+|11\rangle\otimes|11\rangle$. We can
conclude that the bitwise \XOR used here is a better choice of
``multiplication'' in this setting.

Above we used one natural weighting of the bits, while in
information-theoretic considerations often another is natural, namely
equal weighting of the bits for which we obtain
\begin{equation}
  \begin{split}
    &\big|E\big({\cal N}(N_{\alpha,d}\xor N'_{\gamma,d})\big)
    +E\big({\cal N}(N_{\alpha,d}\xor N'_{\delta,d})\big)-d\big|\\
    &\quad+\big|E\big({\cal N}(N_{\beta,d}\xor N'_{\gamma,d})\big)
    -E\big({\cal N}(N_{\beta,d}\xor N'_{\delta,d})\big)\big|
    \leq d,\label{eq:Bell_1}
  \end{split}
\end{equation}
where ${\cal N}(N\xor N')$ is the number of bits set in the bitwise
\XOR of $N$ and $N'$. Again, using the $|\NOPA\rangle$ state,
\begin{equation}
  \begin{split}
    &\big|\langle{\cal N}(n_{0,d}\xor n'_{\gamma,d})\rangle
    +\langle{\cal N}(n_{0,d}\xor n'_{-\gamma,d})\rangle-d\big|\\
    &\quad+\big|\langle{\cal N}(n_{\pitwo,d}\xor n'_{\gamma,d})\rangle
    -\langle{\cal N}(n_{\pitwo,d}\xor n'_{-\gamma,d})\rangle\big|\\
    &\qquad\qquad\qquad\quad =d\left|\cos\gamma\right|
    +\sum_{k=0}^{d-1}K_{2^k}\left|\sin\gamma\right|,
  \end{split}
\end{equation}
and the maximum is obtained as
\begin{equation}
  \sqrt{d^2+\Big(
    \sum_{k=0}^{d-1}K_{2^k}\Big)^2}
  >d.
\end{equation}
At infinite squeezing, the largest possible violation is obtained at
$\sqrt2d$.

\section{Conclusions}
The present paper provides Bell inequalities for number measurements
via the observation that the bits of the number operator are true and
proper qubits. It extends \cite{CPHZ} wherein the authors use the
parity pseudospin system corresponding to the lowest qubit of the
number operator; here we use a commuting hierarchy of similar systems
corresponding to all qubits of $n$. One benefit of this is that the
the available entanglement in the $|\NOPA\rangle$ state is better
used. Furthermore, by this construction, continuous-variable systems
can be used to violate the Bell inequality via number measurements.
Also, via the mentioned qubits, continuous-variable systems are
possible to use for quantum-computational tasks.

Unfortunately, on the experimental side, no good method of measuring
parity is known, much less the extended parity operators $s_{x,d}$,
$s_{y,d}$, and $s_{z,d}$, or the new (truncated) number operators
$n_{x,d}$ and $n_{y,d}$ introduced here. There is an experimentally
challenging proposal in Ref.~\cite{CPHZ} which has not yet been
realized. Nevertheless, perhaps Refs.~\cite{BanaWodk,CPHZ} and this
paper will provide motivation to search for a good measurement
procedure.

Another use of this formalism is to extract distinguishable bipartite
entangled spin systems (or rather, pseudospin systems) out of a
bipartite entangled system where each part consists of
undistinguishable pieces \cite{JKP}. Measuring the \emph{number} of
pieces (atoms or whatnot) that have a certain property instead of
identifying \emph{exactly which} pieces (atoms) that have the property
does seem simpler to achieve. But there are still the experimental
challenges noted above, of course. And it takes $2^k$
undistinguishable pieces of each part of the system to establish $k$
distinguishable spin systems, which can make the procedure
comparatively costly.

Inequalities (\ref{eq:Bell}), (\ref{eq:Bell_d}), (\ref{eq:Bell_B}),
(\ref{eq:Bell_N}), and (\ref{eq:Bell_1}) are all examples of choices
of different weighting of the bits (or, equivalently, the parity
pseudo-spins). In fact, any weighting one finds reasonable can be
used, for example to adapt for the case where the lowest bits are not
really accessible, as may happen in the experimental setup of Ref.\ 
\cite{JKP}.  This is one way of deriving more inequalities from the
above approach, and another is to allow the angles to differ for
different bits. We would then obtain a larger violation from a
finitely squeezed $|\NOPA\rangle$ state than the one shown above,
because the best angles vary from one $d$ to another in the Bell
inequality (\ref{eq:Bell_d}). A third extension is to use qutrits
(spin-1 correspondence) instead of qubits in the approach, or indeed
so-called qu$N$its (spin-$(N-1)/2$ correspondence) for
arbitrary $N$, together with an inequality more suited to such a
situation \cite{KGZMZ,CGLMP}.

As to noise sensitivity and other experimental problems, previous
results are of course usable in a bitwise analysis as indicated
above. But since an error may affect several bits in the
number-measurement approach, a more detailed analysis is necessary.
Further analysis is perhaps of limited value until a good proposal for
measuring $n_x$ and $n_y$ is available, wherein the properties of
possible experimental problems are better visible.

Finally, this is certainly not the only approach to obtain Bell
inequalities for continuous-variable systems, even when taking into
account the possible extensions mentioned above. But the present
treatment is a step towards understanding the difficult and
interesting issues at hand.

\begin{acknowledgments}
  The author would like to thank E.~Polzik and K.~M\o lmer for
  valuable discussions.  This work has been supported by the
  Wenner-Gren Foundation and the Royal Swedish Academy of Science.
  MaPhySto, The Centre for Mathematical Physics and Stochastics, is
  funded by the Danish National Research Foundation.
\end{acknowledgments}

\appendix


\end{document}